\documentclass[twocolumn,prb,showpacs,preprintnumbers,amsmath,amssymb]{revtex4-1}
\usepackage{geometry}
\usepackage{color}
\usepackage{graphicx}% Include figure files \usepackage{dcolumn}
\begin{document}

\title{$d$-wave superconducting gap observed in protect-annealed electron-doped cuprate superconductors Pr$_{1.3-x}$La$_{0.7}$Ce$_{x}$CuO$_{4}$}

\author{M. Horio$^1$\thanks{horio@wyvern.phys.s.u-tokyo.ac.jp}, K. Koshiishi$^1$, S. Nakata$^1$, K. Hagiwara$^1$, Y. Ota$^2$, K. Okazaki$^2$, S. Shin$^2$, S. Ideta$^3$, K. Tanaka$^3$, A. Takahashi$^4$, T. Ohgi$^4$, T. Adachi$^5$, Y. Koike$^4$, and A. Fujimori$^{1,6}$}

\affiliation{$^1$Department of Physics, University of Tokyo, Bunkyo-ku, Tokyo 113-0033, Japan}
\affiliation{$^2$Institute for Solid State Physics (ISSP), University of Tokyo, Kashiwa, Chiba 277-8581, Japan}
\affiliation{$^3$UVSOR Facility, Institute for Molecular Science, Okazaki 444-8585, Japan} 
\affiliation{$^4$Department of Applied Physics, Tohoku University, Sendai 980-8579, Japan} 
\affiliation{$^5$Department of Engineering and Applied Sciences, Sophia University, Tokyo 102-8554, Japan} 
\affiliation{$^6$Department of Applied Physics, Waseda University, Tokyo 169-8555, Japan}

\begin{abstract}
For electron-doped cuprates, the strong suppression of antiferromagnetic spin correlation by efficient reduction annealing by the ``protect-annealing" method leads to superconductivity not only with lower Ce concentrations but also with higher transition temperatures. To reveal the nature of this superconducting state, we have performed angle-resolved photoemission spectroscopy measurements on protect-annealed electron-doped superconductors Pr$_{1.3-x}$La$_{0.7}$Ce$_{x}$CuO$_{4}$ and directly investigated the superconducting gap. The gap was found to be consistent with $d$-wave symmetry, suggesting that strong electron correlation persists and hence antiferromagnetic spin fluctuations remain a candidate that mediates Copper pairing in the protect-annealed electron-doped cuprates.
\end{abstract}

%%% Keywords are not needed any longer. %%%
%%%\kword{keyword1, keyword2, keyword3, \ldots}
%%%

\maketitle
%General
The symmetry of the superconducting (SC) gap provides a clue for the origin of superconductivity. It is now firmly established that the SC gap of the hole-doped cuprate superconductors has $d$-wave symmetry, reflecting the fact that electrons are strongly correlated and tend to avoid double occupation \cite{Scalapino2012}. As for the electron-doped cuprate superconductors with the Nd$_2$CuO$_4$ (so called T'-type) structure, many of previous studies have also supported $d$-wave symmetry \cite{Tsuei2000,Kokales2000,Sato2001,Armitage2001a,Blumberg2002,Zheng2003,Matsui2005b,Dagan2007,Santander2011}. In particular, electronic Raman scattering \cite{Blumberg2002} and angle-resolved photoemission spectroscopy (ARPES) \cite{Matsui2005b} studies have revealed that the SC gap exhibits $d$-wave momentum dependence as well as a maximum near the hot spot, where the antiferromagnetic (AF) Brillouin zone boundary and the Fermi surface cross. Thus, large contribution of AF spin fluctuations to the superconductivity has been proposed though the intrinsic momentum dependence of the SC gap still remains elusive since the coexistence with the AF order would also modulate the SC gap from the monotonic $d$-wave form \cite{Yuan2006}.

%AF correlation vs gap symmetry
The AF correlation in the electron-doped cuprates strongly depends on the post-growth annealing conducted in a reducing atmosphere \cite{Mang2004a,Richard2007}. By the reduction annealing, impurity oxygen atoms at the apical site, which exist in as-grown samples and act as a strong scattering center \cite{Xu1996}, are presumably removed \cite{Radaelli1994,Schultz1996}. Recently, a new annealing method, which is called protect annealing, has been demonstrated to induce superconductivity in electron-doped cuprates with lower Ce concentration and higher $T_c$ \cite{Adachi2013} than those in previous studies \cite{Sun2004}. There, while annealing, single crystals were covered with polycrystalline powders with the same composition, allowing to apply more strongly reducing condition without surface decomposition. With this improved annealing method, the impurity apical oxygen atoms can probably be more efficiently removed. Our ARPES study on the protect-annealed Pr$_{1.3-x}$La$_{0.7}$Ce$_{x}$CuO$_{4}$ (PLCCO, $x = 0.10$) crystals has revealed strong suppression of the hot spot, namely, the AF pseudogap, suggesting a dramatic reduction of the AF spin correlation length and/or the magnitude of the magnetic moments \cite{Horio2016}. A question arising here is what the character of this superconductivity under suppressed antiferromagentism is. Several penetration depth studies \cite{Alff1999,Skinta2002} have reported the observation of $s$-wave SC gap for optimally doped electron-doped cuprate superconductors. Biswas \textit{et al.} \cite{Biswas2002} have claimed, by use of point-contact spectroscopy, that the SC-gap symmetry changes from $d$-wave to $s$-wave in going from underdoped to overdoped regime. These results point toward the possibility of $s$-wave superconductivity in the optimal to overdoped samples where the AF order becomes less relevant. In this context, the SC gap symmetry of protect-annealed samples with suppressed antiferromagnetism is of great interest.

\begin{figure*}[ht!]
\begin{center}
\includegraphics[width=175mm]{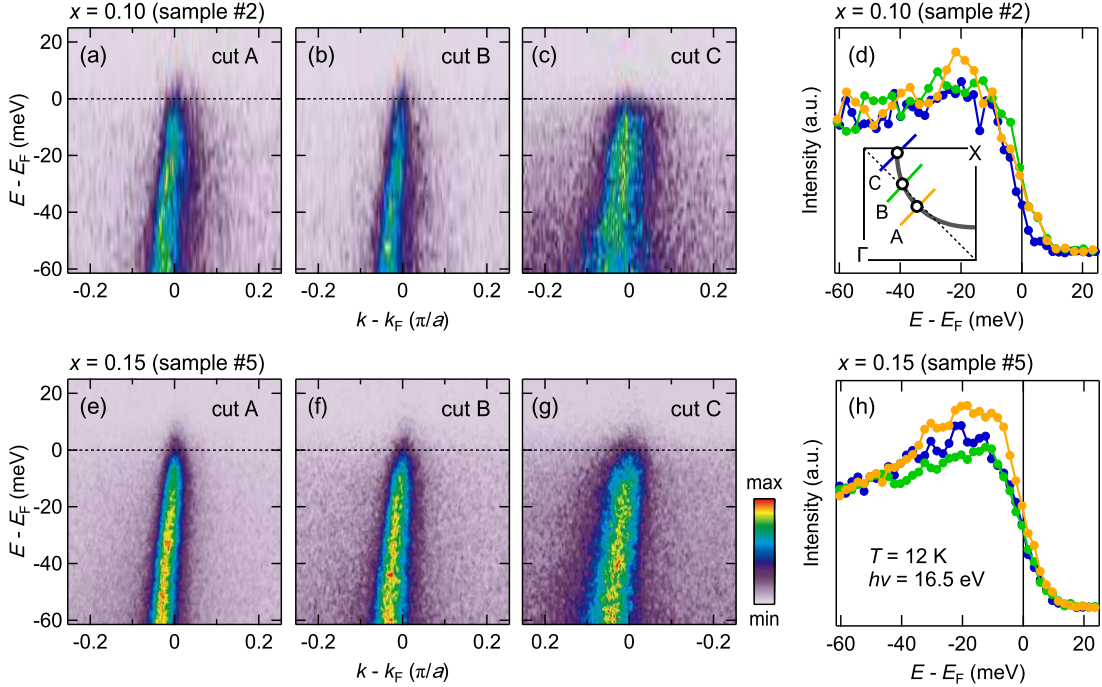}
\end{center}
\caption{ARPES spectra of protect-annealed PLCCO ($x = 0.10, 0.15$) recorded within 1 hour after cleavage. (a)--(c) ARPES spectra of sample \#2 with $x = 0.10$ along cuts A--C indicated in the inset of (d). The spectra were recorded at $T = 12$ K with $h\nu = 16.5$ eV incident photons. (d) EDCs at $k_\mathrm{F}$ points extracted from (a)--(c). The spectra have been normalized to the intensity at the high binding energies of 45--60 meV. (e)--(h) The same as (a)--(d) but for sample \#5 with $x = 0.15$.}
\label{Band}
\end{figure*}

%In this paper
Here, we report on an ARPES study of protect-annealed PLCCO ($x = 0.10, 0.15$) single crystals conducted to reveal the nature of the SC state with the enhanced $T_c$ and suppressed AF correlation. With special care about the surface degradation, we have succeeded in the direct observation of momentum-dependent SC gap. The obtained SC gap suggests its $d$-wave symmetry for both samples, suggesting that AF correlation arising from strong electron correlation remains an essential ingredient for the superconductivity in efficiently annealed electron-doped cuprates.

\begin{figure*}
\begin{center}
\includegraphics[width=175mm]{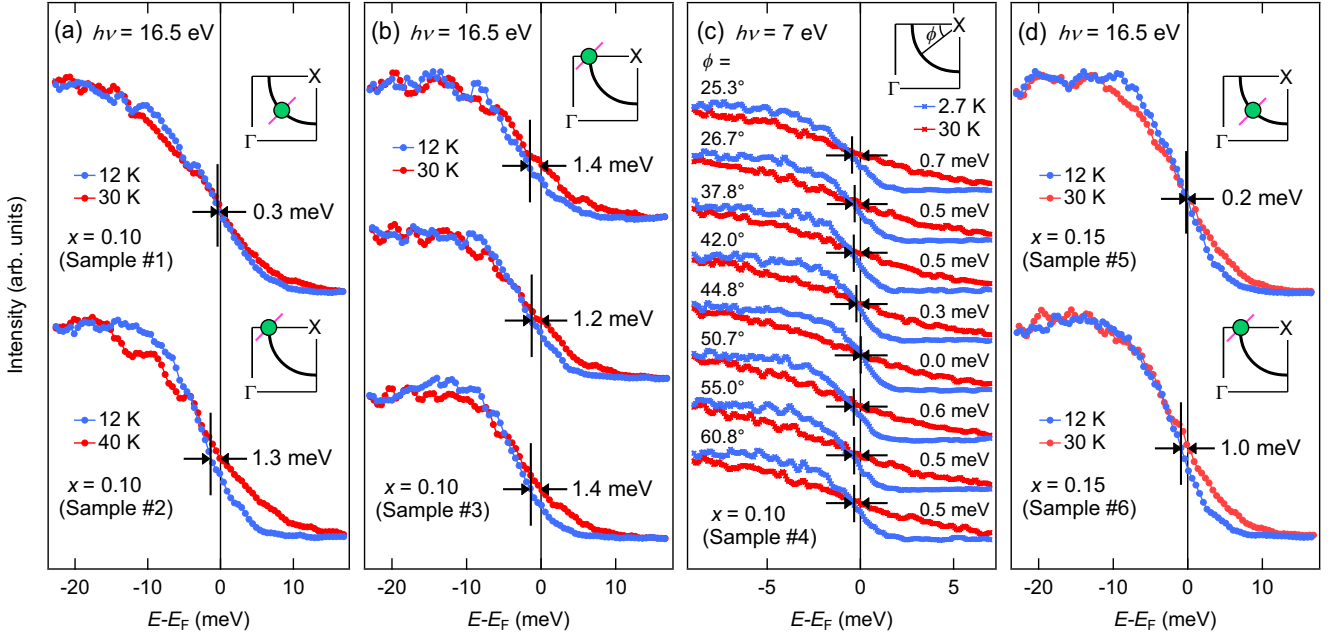}
\end{center}
\caption{Leading-edge shift $\Delta_\mathrm{LE}$ observed for PLCCO samples. EDCs of (a) samples $\#$1 and $\#$2 with $x = 0.10$, (b) sample $\#$3 with $x = 0.10$ measured three times on different cleavage surfaces, (c) sample $\#$4 with $x = 0.10$, and (d) $\#$5 and $\#$6 with $x = 0.15$ measured at temperatures above (red curves) and below (blue curves) $T_c$. Insets of panels (a), (b), and (d) indicate the momentum cuts and $k_{\mathrm{F}}$ positions where the EDCs were measured. For each EDC in panel (c), Fermi surface angle $\phi$ defined in the inset is indicated. Estimated $\Delta_\mathrm{LE}$ values are shown beside each set of the EDCs.}
\label{EDC}
\end{figure*}

%Experimental
Single crystals of PLCCO with $x=0.10$ (samples \#1--\#4) and $x=0.15$ (samples \#5 and \#6) were synthesized by the traveling-solvent floating-zone method and were protect-annealed for 24 hours at 800 $^\circ$C~\cite{Adachi2013}. After the annealing, samples with $x = 0.10$ and $x = 0.15$ showed $T_c$ values of 27 K and 22 K, respectively. Most of the ARPES measurements were performed at beamline 7U of UVSOR-III Synchrotron (samples \#1--\#3, \#5, and \#6). At UVSOR-III, linearly polarized light with $h\nu = 16.5$ eV was used for the measurements. The total energy and momentum resolutions were set at 8 meV and 0.005~\AA$^{-1}$, respectively. Sample $\#3$ was measured three times while each of the other samples was measured once. Prior to each measurement, the sample was cleaved {\it in situ} under the pressure better than 1 $\times$ 10$^{-10}$ Torr. Considering the relatively quick surface degradation of the T'-type cuprates \cite{Sato2001}, ARPES spectra were recorded within 4 hours after cleavage at only one momentum cut at two temperatures below and above $T_c$ for each sample. Just before or after taking every single spectrum of the sample, a gold film evaporated near the sample was measured to determine the Fermi level ($E_\mathrm{F}$) of the sample at that moment. Sample \#4 was measured using a laser-ARPES apparatus developed at the Institute for Solid State Physics (ISSP) with a 7 eV quasi-CW laser (with the repetition rate of 240 MHz). The space-charge effect was confirmed to be negligibly small. The total energy and momentum resolutions were set at 1.5 meV and 0.002~\AA$^{-1}$, respectively. The measurements were carried out in a vacuum better than $4 \times 10^{-11}$ Torr, and several momentum cuts were measured. Energy distribution curves (EDCs) at Fermi momentum ($k_\mathrm{F}$) positions presented in Figs.~\ref{Band} and \ref{EDC} were obtained by integration within $k_\mathrm{F} \pm 0.006 \pi/a$ where $a=3.98$~\AA\ is the in-plane lattice constant.

%Band
Figure~\ref{Band} displays ARPES spectra of protect-annealed PLCCO ($x=0.10, 0.15$) measured at $h\nu = 16.5$ eV near ($\pi/2$, $\pi/2$), hot spot, and (0.3$\pi$, $\pi$). The spectra were recorded within 1 hour after cleavage, when the cleaved surface remained fresh. EDCs were extracted at $k_\mathrm{F}$ points and plotted in Figs.~\ref{Band}(d) and (h) after normalization to the intensity at binding energies of 45--60 meV. For both the $x=0.10$ and $x=0.15$ samples, one cannot recognize remarkable intensity suppression at the hot spot near $E_\mathrm{F}$. This suggests a strong suppression of AF spin correlation length and/or the magnitude of the magnetic moment by protect annealing, consistent with the previous ARPES study on the protect-annealed crystals \cite{Horio2016}.

%EDCs
In order to examine the SC gap of protect-annealed PLCCO samples, in Fig.~\ref{EDC}(a), EDCs of PLCCO ($x = 0.10$) near ($\pi/2$, $\pi/2$) and (0.3$\pi$, $\pi$), i.e., near the node and antinode in the case of $d_{x^2-y^2}$ symmetry are plotted. Because the T'-type cuprates do not show clear SC coherence peaks in their ARPES spectra \cite{Sato2001,Armitage2001a,Matsui2005b} and the position of the leading-edge midpoint at $T < T_c$ referenced to $E_\mathrm{F}$ is neither clear nor necessarily reflecting the magnitude of the SC gap \cite{Armitage2001a}, we compared the spectra taken above and below $T_c$ and estimated the magnitude of the leading-edge shift $\Delta_\mathrm{LE}$ between the two temperatures. Here, $\Delta_\mathrm{LE}$ is defined as an energy shift with decreasing temperature referenced to $E_\mathrm{F}$-crossing point at $T > T_c$. As shown in Fig.~\ref{EDC}(a), the EDCs near ($\pi/2$, $\pi/2$) taken above and below $T_c$ cross almost at $E_\mathrm{F}$ and $\Delta_\mathrm{LE}$ is as small as 0.3 meV, whereas those near (0.3$\pi$, $\pi$) cross appreciably below $E_\mathrm{F}$ and a leading-edge shift of 1.3 meV was observed. The leading-edge shift of similar magnitude near (0.3$\pi$, $\pi$) was also observed in the measurements of another $x = 0.10$ sample reproducibly on different cleavage surfaces as shown in Fig.~\ref{EDC}(b), clearly indicating the SC gap opening near (0.3$\pi$, $\pi$). The momentum region near (0.3$\pi$, $\pi$) cannot be reached in the ARPES measurement using 7 eV laser due to the low photon energy, but the laser-ARPES spectra also show negligibly small $\Delta_\mathrm{LE}$ near ($\pi/2$, $\pi/2$) and finite $\Delta_\mathrm{LE}$ away from ($\pi/2$, $\pi/2$) [Fig.~\ref{EDC}(c)]. The same tendency is also observed for the $x = 0.15$ samples [Fig.~\ref{EDC}(d)], although $\Delta_\mathrm{LE}$ near (0.3$\pi$, $\pi$) was slightly smaller than that of the $x = 0.10$ samples.

\begin{figure}
\begin{center}
\includegraphics[width=90mm]{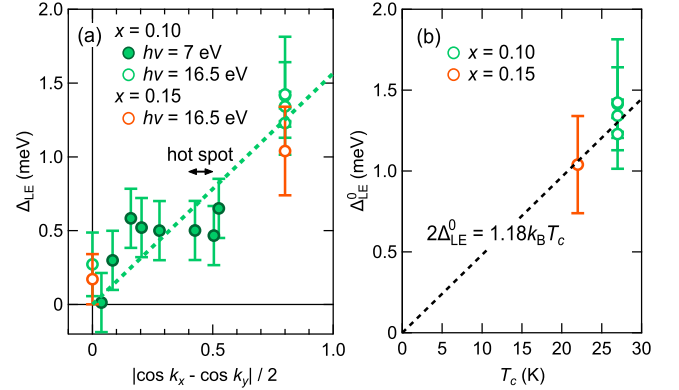}
\end{center}
\caption{$d$-wave SC gap of PLCCO. (a) Leading-edge shift $\Delta_\mathrm{LE}$ plotted against the $d_{x^2-y^2}$-wave order parameter $|\mathrm{cos} k_x - \mathrm{cos} k_y| / 2$. A line obtained by fitting the data is also plotted. (b) $\Delta_\mathrm{LE}$ at the antinode (0.3$\pi$, $\pi$), $\Delta_\mathrm{LE}^0$, plotted against $T_c$ values. A line representing $2\Delta_\mathrm{LE}^0 = 1.18k_\mathrm{B}T_c$ is superimposed.}
\label{Gap}
\end{figure}

\begin{figure*}[ht!]
\begin{center}
\includegraphics[width=140mm]{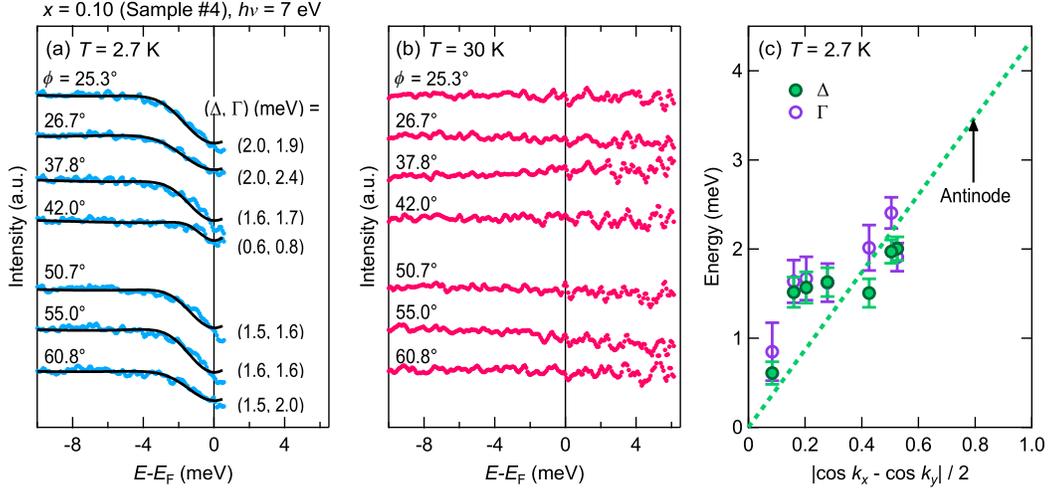}
\end{center}
\caption{SC gap estimated from TDOS. (a),(b) Off-nodal TDOS of the $x = 0.10$ sample derived from laser-ARPES data at $T = 2.7$~K and 30~K, respectively. The $T = 2.7$~K spectra are fitted to the Dynes formula (shown in black). (c) SC gap $\Delta$ and pair-breaking scattering rate $\Gamma$ obtained through fitting and plotted against the $d_{x^2-y^2}$-wave order parameter $|\mathrm{cos} k_x - \mathrm{cos} k_y| / 2$.}
\label{TDOS}
\end{figure*}

%Gap plot%
The $\Delta_\mathrm{LE}$ values estimated from Fig.~\ref{EDC} are plotted against the $d_{x^2-y^2}$-wave order parameter $|\mathrm{cos} k_x - \mathrm{cos} k_y| / 2$ in Fig.~\ref{Gap}(a). A tiny but finite drift of the incident photon energy during the synchrotron measurement, which affects the kinetic energy of the photoelectrons, was evaluated from the drift of $E_\mathrm{F}$ of the reference gold film and indicated by an error bar, while the error bars for the laser-ARPES data were assumed to be constant. The $\Delta_\mathrm{LE}$ for the $x = 0.10$ samples is roughly proportional to the $d$-wave order parameter. For more detailed discussion about the momentum dependence of the SC gap such as the deviation from the monotonic $d_{x^2-y^2}$-wave form \cite{Blumberg2002,Matsui2005b}, more thorough investigation especially around the hot spot is required. Still, at this moment, one can conclude that the observed SC gap is consistent with $d$-wave symmetry. Detection of the sign change using phase-sensitive probes is a future issue. In Fig.~\ref{Gap}(b), $\Delta_\mathrm{LE}$ at $\sim (0.3\pi, \pi)$ is plotted as $\Delta_\mathrm{LE}^0$ against $T_c$. The dependence of the antinodal $\Delta_\mathrm{LE}$ on $T_c$ has been satisfactorily fitted to the equation of $2\Delta_\mathrm{LE}^0 = \alpha k_\mathrm{B}T_c$. This observation suggests that the SC states of $x = 0.10$ and $x = 0.15$ samples are in the same $d$-wave symmetry and are realized under the same mechanism. On the other hand, obtained value of $\alpha = 1.18$, which represents the paring strength, is quite small even compared to $\alpha = 2\Delta / k_\mathrm{B}T_c = 4.28$ predicted by $d$-wave BCS theory. In fact, it has been shown in previous ARPES studies that the leading-edge shift gives an underestimate of the SC gap magnitude by a factor of $\sim 2$ \cite{Kondo2007,Yoshida2009,Reber2012}.

%TDOS
In order to gain more quantitative information, we attempt to apply the recently developed tomographic density of states (TDOS) method \cite{Reber2012}. TDOS is defined as sum of EDCs along one momentum cut which is normalized to similar but ungapped referenced sum along the nodal direction. The resulting TDOS can be fitted to the following Dynes formula:

\begin{equation}
I_\mathrm{TDOS} (\omega) = \mathrm{Re} \frac{\omega - i \Gamma}{\sqrt{(\omega - i \Gamma)^2 - \Delta^2}}
\end{equation}

where $I_\mathrm{TDOS}$ is the TDOS intensity, $\Gamma$ is the pair-breaking scattering rate, and $\Delta$ is the SC gap~\cite{Dynes1978}. Without further complicated assumptions about fitting, this method is capable of determining SC gap magnitude more precisely than conventional estimates from leading-edge shifts or symmetrized EDCs \cite{Reber2012}. Figures~\ref{TDOS}(a) and (b) show TDOS obtained from the laser-ARPES measurement on the $x = 0.10$ sample [same data as in Fig.~\ref{EDC}(c)]. While clear gaps are found at all the off-nodal positions at $T = 2.7$~K [Fig.~\ref{TDOS}(a)], the gaps are closed at $T = 30$~K $> T_c = 27$~K [Fig.~\ref{TDOS}(b)], consistent with $d$-wave SC gap opening below $T_c$. SC gap value $\Delta$, obtained through fitting to the Dynes formula, is plotted in Fig.~\ref{TDOS}(c) as a function of the $d_{x^2-y^2}$-wave order parameter $|\mathrm{cos} k_x - \mathrm{cos} k_y| / 2$. The overall trend of the momentum dependence is consistent with that of the leading-edge shift shown in Fig.~\ref{Gap}(a). As $\Delta$ increases, the pair-scattering rate $\Gamma$ also increases, thereby suppressing the coherence peak. From linear extrapolation of the obtained $\Delta$ values, antinodal SC gap value at (0.3$\pi$, $\pi$) ($|\mathrm{cos} k_x - \mathrm{cos} k_y| / 2 \sim 0.8$) can be estimated to be 3.5~meV. This leads to $\alpha = 2\Delta / k_\mathrm{B}T_c = 3.0$, which is $\sim 2.5$ times larger than the estimate from leading-edge shift. Still, the value of $\alpha$ is somewhat smaller than what is predicted by $d$-wave BCS theory, and it remains elusive whether this relatively small $\alpha$ is intrinsic or due to an artifact of the analysis. With any analysis methods, the exact size of the SC gap cannot be uniquely determined when the SC coherent peak is not clear as in the present case.

Studies on protect-annealed PLCCO samples using muon spin relaxation \cite{Adachi2016} and NMR \cite{Yamamoto2016}, which are sensitive to spins, have revealed that while the long-range AF order is suppressed, the AF spin susceptibility increases with decreasing temperature for both $x = 0.10$ and $x = 0.15$ samples, and short-ranged AF order sets in at very low temperatures for the $x = 0.10$ sample. Given the presence of the enhanced AF spin susceptibility, a simple but straightforward scenario is to associate the $d$-wave superconductivity suggested from the present ARPES results with AF spin fluctuations that arise from strong electron correlation. This consideration is consistent with a phenomenological model in which the T'-type cuprates are regarded as being in an antiferromagnetically correlated state and a static short-range AF order is induced around the excess apical oxygen atoms \cite{Adachi2013,Adachi2016}. Considering the fact that the efficient removal of the apical oxygen atoms by protect annealing results in the increase of $T_c$ compared to the conventional annealing \cite{Sun2004}, the short-range AF order induced around the apical oxygen atoms is harmful for superconductivity. Once the apical oxygen atoms are removed, AF correlation becomes more short ranged to nurture superconductivity. Emergence of the superconductivity and its relationship to the AF pseudogap have been discussed by ARPES studies in terms of the reduction of spin correlation length~\cite{Park2013,Horio2016,Song2017}. Thus, strong electron correlation leading to AF spin fluctuations remains a candidate that drives the $d$-wave superconductivity in the T'-type cuprates.

%Conclusion%
In conclusion, we have performed ARPES measurements on protect-annealed PLCCO single crystals ($x = 0.10, 0.15$) and estimated the leading-edge shift $\Delta_\mathrm{LE}$ as a measure of the SC gap. Observed momentum dependence of $\Delta_\mathrm{LE}$ was consistent with $d$-wave symmetry, suggesting that superconductivity in the T'-type cuprates may be driven by AF spin fluctuations arising from strong electron correlation regardless of the doping level even after the suppression of the static short-range AF order and the strong reduction of AF spin correlation length and/or the magnitude of the magnetic moment by protect annealing.

\begin{acknowledgments}
ARPES experiments were performed at UVSOR-III Synchrotron (Proposal Nos. 27-808 and 28-545) and ISSP. This work was supported by JSPS KAKENHI Grants No.~JP14J09200, No.~JP15H02109, No.~JP16K05458, No.~JP17H02915, No.~JP19H00651, No.~JP19H01818, No.~JP19H01841, No.~JP19H05826, and No.~JP19K03741. M.H. acknowledges support from the Advanced Leading Graduate Course for Photon Science (ALPS) and the JSPS Research Fellowship for Young Scientists.
\end{acknowledgments}

%\bibliography{reference}

\end{document}